\documentclass[twocolumn,showpacs,preprintnumbers,amsmath,amssymb,prl]{revtex4}

\usepackage{graphicx}
\usepackage{dcolumn}
\usepackage{bm}
\usepackage{latexsym}
\usepackage{amstext}
\usepackage{amsxtra}

\newcommand{\bs}{\boldsymbol}

\begin{document}
\title{Dynamics of a single vortex line in a Bose-Einstein condensate}
\author{P. Rosenbusch, V. Bretin, and J. Dalibard}
\affiliation{Laboratoire Kastler Brossel$^*$, 24 rue Lhomond,
75005 Paris, France}
\date{Received June 26, 2002}

\begin{abstract}
{We study experimentally the line of a single quantized vortex in
a rotating prolate Bose-Einstein condensate confined by a harmonic
potential. In agreement with predictions, we find that the vortex
line is in most cases curved at the ends. We monitor the vortex
line leaving the condensate. Its length is measured as a function
of time and temperature. For a low temperature, the survival time
can be as large as 10 seconds. The length of the line and its
deviation from the center of the trap are related to the angular
momentum per particle along the condensate axis.}
\end{abstract}


\maketitle

The macroscopic motions of quantum and classical fluids are
dramatically different. The description of a quantum fluid by a
macroscopic wave function $\psi=\sqrt{\rho}\; e^{i\theta}$ imposes
strong constraints upon its velocity field. At a position of
non-vanishing density $\rho$, the velocity is related to the phase
$\theta$ by $\bs v=\hbar\,\bs \nabla\theta/m$, where $m$ is the
mass of a particle of the fluid. Hence $\bs \nabla \times \bs
v=0$. A rotational motion of the fluid can be obtained only
through the nucleation of vortex lines, along which the density is
zero and around which the circulation of the velocity is quantized
in units of $h/m$ \cite{Onsager,Feynman}.

Quantized vortices play an essential role in the dynamics of all
quantum macroscopic objects. Examples are flux lines in
superconductors \cite{Tinkham}, and vortex lines in superfluid
liquid helium \cite{Donnelly91} and gaseous Bose-Einstein
condensates (BEC)
\cite{Cornellphaseimprinting,Madison00,Cornellcooling,Ketterle,Hodby01}.
Among the remaining problems, is the shape of a vortex/flux line
and the study of its time evolution. The observation of inclined
flux lines in superconductors was only possible due to recent
advances in electron microscopy \cite{fluxlineMFM}. In gaseous
BEC, one has comparably easy access to the vortex line because the
density of the atom cloud is low. A few disordered vortex lines
have been observed by taking tomographic images perpendicular to
the long axis of a cigar shaped condensate \cite{Ketterle}. An
array of many vortex lines in a pancake shaped condensate has been
observed by transverse imaging of the whole atom cloud
\cite{Engels02}.

In this Letter we report the full length observation of a single
vortex line in a cigar shaped condensate.  We find that as a
result of spontaneous symmetry breaking the line is generally
bent. Our experimental results confirm recent predictions, in
which the shape of the vortex line minimizing the energy of the
gas was derived for a given rotation frequency
\cite{Garcia01a,Garcia01b,Aftalion01,Modugno,Aftalion02}. We also
study the time evolution of the shape of the line. As the angular
momentum of the gas slowly decays, the bending of the vortex line
and its deviation from to the center of the trap increase. We
investigate the influence of temperature on these dynamics. From
our results one can hope to draw some indications for the shape
and dynamics of the vortex/flux line in systems where a direct
observation is not yet possible.

Our $^{87}$Rb condensate is formed by radio-frequency (rf)
evaporation of $10^9$ atoms in a Ioffe-Pritchard magnetic trap.
The atoms are spin-polarized in the $F=m_F=2$ state. The magnetic
trap has a longitudinal frequency $\omega_z/2\pi=11.8$~Hz and a
transverse frequency
$(\omega_x+\omega_y)/4\pi=\omega_\bot/2\pi=97.3$~Hz (the $x$ axis
is vertical). Because gravity slightly displaces the center of the
trap with respect to the magnetic field minimum, the potential in
the $xy$ plane is not perfectly isotropic and we measure a 1\%
relative difference between $\omega_x$ and $\omega_y$.

The initial temperature of the cloud, which is pre-cooled using
optical molasses, is 100 $\mu$K. The condensation threshold is
reached at $T_{\rm c}\sim300$~nK, with $N_{\rm
c}\sim2\;10^6$~atoms. We cool to typically $T\sim 90$~nK which
corresponds to a condensate with $N_0\sim5\;10^5$~atoms and a
chemical potential $\mu\sim 70$~nK. It is obtained using $\nu_{\rm
f}=\nu_0 +10$~kHz as the final rf, where $\nu_0$ is the frequency
at which the trap is emptied. During the rest of the experimental
cycle, we maintain the evaporation rf at an adjustable level,
typically $\nu=\nu_0 +12$~kHz. This allows us to control the
temperature while observing the vortex line.

Once the condensate is formed, we use an off-resonant laser beam
to impose on the trapping potential an elliptic anisotropy in the
$xy$ plane \cite{Madison00}. The wavelength of the beam is 852~nm,
its power 0.1~mW and its waist $20\;\mu$m. Acousto-optic
modulators deflect the position of the beam in the $xy$ plane,
thereby rotating the potential anisotropy at a frequency of
$\Omega/2\pi=70$~Hz. We apply this ``laser stirrer" for 300~ms,
during which 7 or more vortices are nucleated \cite{Madison01}.
After the stirring phase, the condensate evolves freely in the
magnetic trap for an adjustable time $\tau$.

The preparation of a single vortex line takes advantage of the
slight static anisotropy of our magnetic trap, so that the angular
momentum is not exactly a constant of motion. In a time $\tau \sim
1$--$2$~s, we observe a transition from a multi-vortex condensate
to a condensate with a single vortex. This relatively long time
levels the fluctuations that may occur during the nucleation
process. Thereby we are able to reproduce a single vortex
condensate on every experimental cycle. This vortex line can then
be studied for a time $\tau \leq 10$~s.

\begin{figure}
 \centerline{\includegraphics{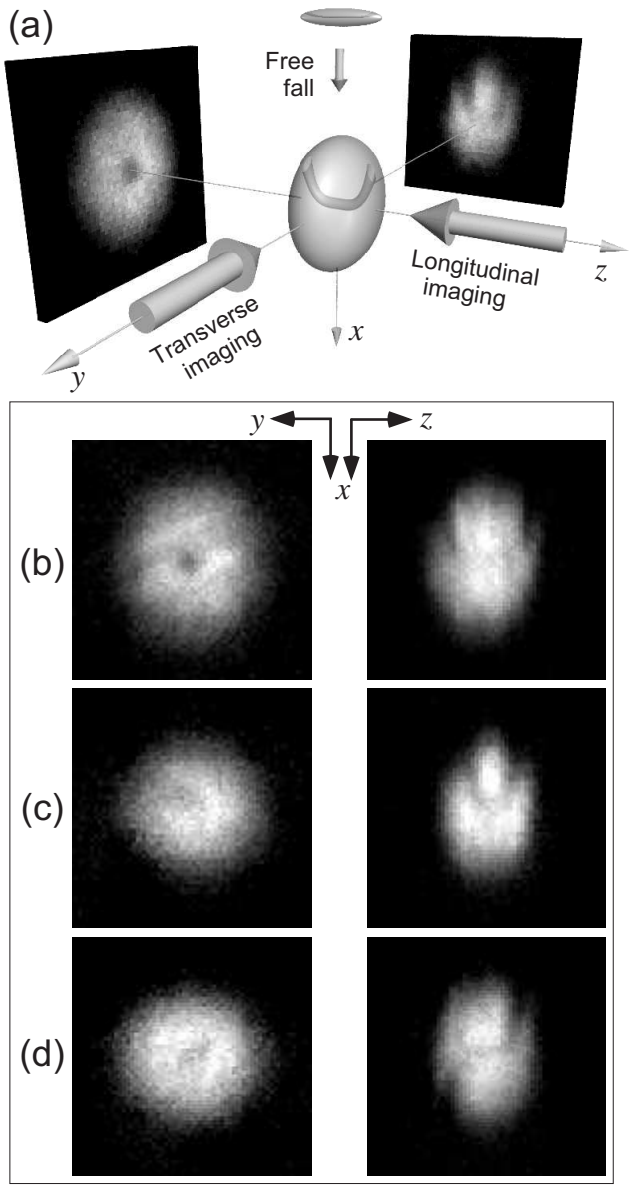}}
\caption{(a) Schematic of the imaging system. The cigar shaped
condensate is imaged after 25~ms of time of flight leading to the
inversion of the ellipticity in the $xz$ plane. Two beams image
the atom cloud simultaneously along the longitudinal ($z$) and
transverse ($y$) directions of the initial cigar. (b-d)
Simultaneous longitudinal and transverse images of condensates
after $\tau=4$~s (b), $\tau=7.5$~s (c) and $\tau=5$~s (d).}
\label{fig:images}
\end{figure}

The atom distribution at $\tau$ is probed (destructively) by
switching off the magnetic trap, letting the cloud expand during
$t_{\rm TOF}=25$~ms and performing absorption imaging. Two imaging
beams aligned along the $y$ and $z$ directions probe the atom
cloud simultaneously (Fig.~\ref{fig:images}a). The beams are
combined onto a camera with the same magnification. During the
expansion, the transverse dimensions $x$ and $y$ of the condensate
are magnified by $\omega_\bot t_{\rm TOF}\sim 15$, while the
longitudinal dimension is nearly unchanged \cite{Castin96}. It has
been shown theoretically that the presence of a single vortex line
does not alter this expansion and that the coordinates of the line
are scaled by the same factors \cite{Modugno}.

Figs.~\ref{fig:images} b-d show three condensates imaged after
various times. The left column shows the ``longitudinal" view
along $z$, representing the atom distribution in the $xy$ plane.
The right column depicts the ``transverse" view taken along the
$y$ direction, representing the atom distribution in the $xz$
plane. The vertical ($x$) direction is identical for all images.
The transverse images show the typical atom distribution where the
ellipticity is inverted with respect to the {\it in situ} cigar
form, caused by the transverse expansion during $t_{\rm TOF}$.

As in \cite{Madison00}, we use the longitudinal images in
Fig.~\ref{fig:images} to verify the presence of a single vortex.
The transverse image in Fig.~\ref{fig:images}b obtained for
$\tau=4$~s shows the vortex line as a line of lower atom density.
Clearly this vortex line is not straight. It rather has the shape
of a wide ``U". The $x$ position of the axial part of the vortex
line (bottom of the U) is close to the center of the condensate.
It coincides with the $x$ position of the dip in density seen in
the longitudinal image. In the laboratory frame, the curved vortex
line is expected to rotate around the $z$ axis with a frequency
related to the angular momentum of the condensate. This is
confirmed by the fact that we observe up- and downwards bending
with equal probability.

Fig.~\ref{fig:images}c shows images taken after $\tau=7.5$~s, for
which the angular momentum has decreased significantly compared to
Fig.~\ref{fig:images}b. In the longitudinal view, one sees a
vortex off-center and in the transverse view a narrow U, where the
bottom of the U no longer extends to the center of the condensate.

Fig.~\ref{fig:images}d shows a vortex line in the shape of an
unfolded ``N", observed after $\tau=5$~s. The width of the N,
which is the projection of the vortex line onto the $z$ axis, is
comparable to the width of the U in Fig.~\ref{fig:images}b. The
fact that U as well as N shaped vortex lines are observed leads to
the question whether the bending occurs in one plane or whether
three dimensional deformations of the vortex line can also occur.
We have indications that this may be the case: some transverse
views (not shown here) reveal asymmetric U or N shaped vortex
lines.

\begin{figure}
\scalebox{0.995}{\centerline{\includegraphics{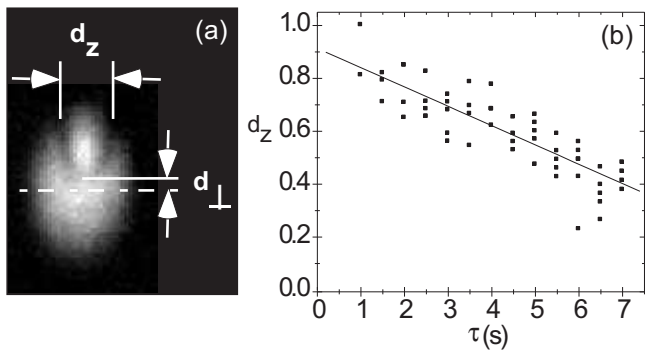}}}
 \vskip -0.3cm
\caption{(a) Schematic of the extraction of $d_z$ and $d_\bot$
from the vortex line. (b) Evolution of $d_z$ with $\tau$. Each
point corresponds to a single image.}
 \label{fig:reproducibility}
\end{figure}

In order to give a quantitative analysis, we measure by hand the
distance along the $z$ direction between the two points where the
vortex line leaves the condensate
(Fig.~\ref{fig:reproducibility}a). We do not distinguish U and N
shaped vortices. Normalization by the length of the condensate
along $z$ leads to the quantity $d_z$.  In
Fig.~\ref{fig:reproducibility}b, we plot $d_z$ as a function of
$\tau$. Each point in the plot corresponds to a single image. The
small spread of the points demonstrates the reproducibility of the
vortex shape. The normalized length of the vortex line $d_z$
decreases quasi-linearly with time.

Fig.~\ref{fig:reproducibility}b shows that after $\tau=7$~s the
vortex is still present. This differs from our earlier reports
\cite{Madison00} simply because we have used two different methods
of identifying the vortex. In \cite{Madison00} only the
longitudinal view was available. Now we find that a vortex can
still be identified in the transverse view, while in the
longitudinal image the density variation due to the bent vortex
line is within the background fluctuations. This long lifetime can
be compared with the MIT result \cite{Ketterle}, where a vortex
array with more than 100 vortices was produced at $\tau=0$. The
number of vortices was divided by $4$ in $\sim 5$~s; however a
single vortex could still be detected after $\tau=40$~s. In both
experiments, it is clear that the decay time of the last vortex is
much longer than the decay time of the initial vortex lattice.

\begin{figure}
\centerline{\includegraphics{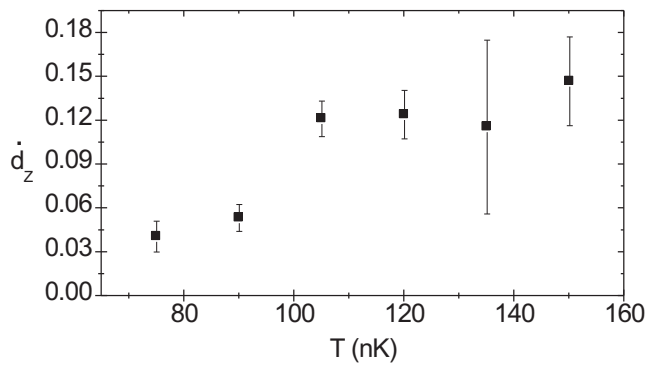}}
 \vskip -0.3cm
 \caption{Variation of $\dot d_z$ with the temperature $T$.
$\dot d_z$ is derived from fits to data similar to those in
Fig.~\ref{fig:reproducibility}b.}
 \label{fig:variationwithT}
\end{figure}

We have repeated the experiment of Fig.~\ref{fig:reproducibility}
for different evaporation radiofrequencies $\nu$ (implying
different temperatures $T$) during the free evolution time $\tau$.
We always observe a quasi-linear decrease of $d_z$ and fit its
slope, $\dot d_z$. It is plotted in Fig.~\ref{fig:variationwithT}
as a function of $T$. Clearly the longest vortex lifetime is
obtained for small temperatures, confirming the idea that the
vortex line is dragged to the edge of the condensate by a
thermally activated process \cite{Fedichev}. Note that at the
lowest temperature we observe almost exclusively U shaped
vortices, while U and N are equally probable at the second lowest
temperature.

In a last experiment we vary again $\tau$ in order to prepare
condensates with various lengths of the vortex line. We measure
the angular momentum per particle $L_z$ by the same method as in
\cite{lz}. The presence of a vortex leads to the lift of
degeneracy between the two quadrupole surface modes carrying
angular momentum $\pm2\hbar$ \cite{Stringari}:
$\omega_{+}-\omega_{-}=2\langle
L_z\rangle/(m\langle{r_\bot}^2\rangle)$, where $r_\bot$ is the
radius in the transverse plane. We excite a superposition of these
two modes by a 1.9~ms flash of our laser stirrer and we probe the
quadrupole oscillation at $t_{\rm osc}=2$~ms and $t_{\rm
osc}=9$~ms after the flash. We repeat the experiment a third time
without the laser flash and analyze the shape of the vortex line.
As above we extract the normalized length $d_z$ of the vortex line
along the $z$ axis. We also measure the displacement $d_\bot$ of
the axial part of the vortex line (bottom of the U) from the
center of the condensate, and normalize it by the radius of the
condensate in the $xy$ plane (Fig.~\ref{fig:reproducibility}a).
Since we have access only to the projection of the decentering on
the $xz$ plane, we actually measure $d_\bot|\cos \alpha|$, where
$\alpha$ is the azimuthal angle of the axial part of the line. We
account for this geometrical factor by dividing the measured
displacement by $\langle |\cos \alpha|\rangle=2/\pi$.

\begin{figure}
 \centerline{\includegraphics{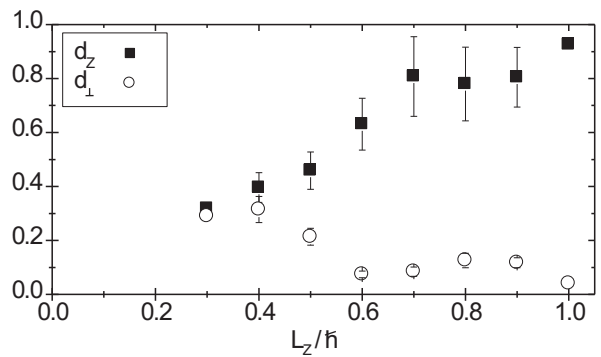}}
 \vskip -0.3cm
\caption{Variation of $d_z$ and $d_\bot$ as functions of the
average angular momentum per particle $L_z$. All measurements were
binned into intervals of $L_z=0.1\;\hbar$ and averaged. The error
bars give the statistical spread.}
 \label{fig:Lz}
\end{figure}

Fig.~\ref{fig:Lz} shows $d_z$ and $d_\bot$ as functions of
$L_z/\hbar$. For clarity we group all points into bins of
$L_z=0.1\;\hbar$ and average over $d_z$ and $d_\bot$. The error
bars give the statistical variation. The data corresponding to
$L_z\leq0.2\;\hbar$ are not reproducible enough and are omitted.
The graph shows that $d_z$ and $L_z/\hbar$ are approximately
equal. A straight ($d_z\sim 1$) and well centered ($d_\bot\ll 1$)
vortex line corresponds to an angular momentum of the order of
$\hbar$. When the angular momentum decreases, we measure a
decentering $d_\bot\leq 0.15$ as long as $L_z>0.5\;\hbar$. Below
$L_z=0.5\;\hbar$, $d_\bot$ rises to 0.3.

We now compare our experimental results with recent predictions
for the shape of a vortex line in an inhomogeneous cigar shaped
condensate
\cite{Garcia01a,Garcia01b,Aftalion01,Modugno,Aftalion02}. These
theoretical studies consist in looking for the ground state of the
condensate in a frame rotating at an angular frequency $\Omega$.
The general conclusion is that above a critical frequency
$\Omega_{\rm c}$ the ground state of the system has one or several
vortices. The central vortex is generally bent if the trap aspect
ratio $\omega_\bot/\omega_z$ is large compared to 1, which is the
case in our experiment. A simple physical picture of the bending
is given in \cite{Modugno}. A cigar shaped condensate can be
viewed as a series of 2D sheets of condensate at various $z$. For
each sheet, one has a 2D vortex problem leading to a critical
frequency $\Omega^{\rm (2D)}_{\rm c}(z)$ above which a centered
vortex is the stable solution. For a given rotation frequency
$\Omega$, it can happen that a centered vortex is the minimal
energy configuration for the sheets close to $z=0$ (i.e.
$\Omega^{\rm (2D)}_{\rm c}(0)<\Omega$), while it is not the case
for the sheets close to the edges of the condensate where the atom
density is lower. In this case, the vortex line minimizing the
total energy is well centered for $|z|<z_{\rm c}$, and is strongly
bent for $|z|>z_{\rm c}$, where $\Omega^{\rm (2D)}_{\rm c}(z_{\rm
c})=\Omega$. A precursor of this bending effect can also been
found in \cite{Svidzinsky,Feder} in which a stability analysis of
a straight vortex in an elongated condensate showed that some
bending modes have negative energy, and are thus unstable.

Our experimental procedure is somewhat different from the one
considered in these theoretical studies. In our case no rotating
anisotropy is imposed onto the condensate during the relevant
evolution. The stirring laser has been switched on for a short
time only, at the beginning of the procedure, in order to set a
non-zero angular momentum in the system. We observe the evolution
of the condensate in our static trap, as the angular momentum of
the gas slowly decays. However, the shape of the vortex line that
we observe at short time in Fig.~\ref{fig:images}b is remarkably
similar to those predicted and plotted in
\cite{Garcia01a,Garcia01b,Aftalion01,Modugno}. As pointed out in
\cite{Garcia01b}, this bending is a symmetry breaking effect which
does not depend on the presence of a rotating anisotropy and which
happens even in a completely symmetric setup \cite{anisotropy}. A
decentered single vortex similar to the one shown in
Fig.~\ref{fig:images}c was also found as the ground state of the
rotating system in \cite{Aftalion01} for a given rotation range.
By contrast, we did not find in the literature predictions for
N-shaped vortices, such as the one shown in
Fig.~\ref{fig:images}d. This probably means that a N-shaped vortex
is slightly more energetic than a U shaped vortex with the same
angular momentum, and that it could not emerge from a procedure
aiming to find the ground state of the system.

Our results also provide information on the dynamics of the vortex
line and the way it escapes from the condensate. A theoretical
model has been proposed, in which the decay is due to the coupling
with the non-rotating thermal component \cite{Fedichev}. In this
model, the authors considered a spatially homogenous condensate in
a cylindrical vessel. The vortex line, assumed to be straight,
spirals out of the condensate. The bending of the line that we
observe experimentally may change qualitatively the picture, since
we find that the decay occurs first by an increased bending and,
only in a second step, by a deviation of the center of the line
(bottom of the U) from the center of the condensate.

In conclusion we have reported the observation of the full line of
a single quantized vortex. We have related its shape (bending and
deviation from center) to the angular momentum of the system. Our
results should help modelling the dissipative evolution of a
rotating Bose-Einstein condensate.

{\acknowledgments We thank Y. Castin, F. Chevy and G. Shlyapnikov
for useful discussions. P. R. acknowledges support by the EC
(contract number HPMF CT 2000 00830). This work was partially
supported by CNRS, Coll\`{e}ge de France, R\'egion Ile de France,
DGA, DRED and EC (TMR network ERB FMRX-CT96-0002). }


\begin{thebibliography}{99}
\bibitem{anisotropy}
The small static anisotropy of our trap should not influence the
symmetry breaking which occurs in the rotating frame.
\end{thebibliography}
\end{document}